\documentclass[pre,twocolumn,aps,eqsecnum]{revtex4}
\usepackage{epsfig}

\begin{document}

\title{Thermal Effects in Dislocation Theory II: Shear Banding and Yielding Transitions}

\author{J.S. Langer}
\affiliation{Department of Physics, University of California, Santa Barbara, CA  93106-9530}

\date{\today}

\begin{abstract}
The thermodynamic dislocation theory presented in preceding papers is used here to describe   shear-banding instabilities.  Central ingredients of the theory are a thermodynamically defined effective configurational temperature, and a formula for the plastic deformation rate determined by thermally activated depinning of entangled dislocations. An important feature of this paper is an interpretation of yielding transitions in polycrystalline solids.     
\end{abstract}

\maketitle

\section{Introduction}
\label{Intro} 

In the preceding paper \cite{JSL-16}, I reviewed basic features of a thermodynamic theory of dislocation-mediated plasticity in polycrystalline solids. I showed there, in an oversimplified toy model, how this theory might explain shear-banding instabilities in such materials.  My purpose here is to use those ideas in a more realistic analysis of shear-banding dynamics.  More generally, I want to explore the implications of this theory in other nonequilibrium situations, especially yielding transitions.  

There is a large body of literature, extending over more than three decades, devoted to what is known as adiabatic shear banding (ASB) in metals and alloys. For example, see \cite{MARCHAND-DUFFY-88, WRIGHT-02, ASL-12}. This subject is important; the banding instability is generally recognized as a principal failure mechanism in rapidly stressed structural materials.  However, the experimental observations of ASB that I have found so far are inadequate for my purposes.  

The thermodynamic dislocation theory described in \cite{LBL-10,JSL-15,JSL-16} has focussed on strain hardening and related strain-rate dependent phenomena. It describes those phenomena in terms of a small number of physically meaningful state variables that are consistent with basic principles of nonequilibrium statistical physics \cite{BL-I-09}. The ASB observations, however, are generally not accompanied by measurements that allow me to determine equations of motion for those variables.  When stress-strain curves are shown in the literature, they usually show a yielding transition at a large stress and a very small strain, and then a sudden stress drop at a larger strain indicating failure.  As will become clear here, that initial yielding transition is strongly sensitive to sample preparation.  It cannot tell us much about the intrinsic dynamical properties of the material.  

My main theme in this paper is that ASB is a remarkably deep probe of the internal dynamics of  structural materials.  The ``adiabaticity'' of ASB refers to the idea that these banding instabilities are caused by thermal softening in situations where heat flow is slower than plastic deformation.  A local increase in strain rate produces a local increase in heat generation that, in turn, softens the material and further increases the local strain rate.  The result is a runaway instability if the heat is unable to flow away from the hot spot more quickly than new heat is being generated there. Thus, we are looking at a delicate balance between thermal and mechanical behaviors.  To understand what is happening, we need a first-principles theory of the underlying deformation mechanism.  I cannot find the information that I need for developing such a theory in the existing ASB literature. 

To work around this difficulty, I will use the same strategy here that I used in \cite{JSL-16}.  Thanks to the pioneering work of Kocks, Mecking, Follansbee, Meyers and others \cite{KOCKS-MECKING-03,FOLLANSBEE-KOCKS-88,MEYERSetal-95}, we have a first-principles picture of plastic deformation in copper. (Other papers that I have found useful for understanding the present state of this field include \cite{MEYERSetal-06,ARMSTRONGetal-09,GRAY-12}.) The trouble is that copper is not observed to undergo ASB, probably because its thermal conductivity is too high.  In \cite{JSL-16}, I invented a ``pseudo copper'' by using the material parameters that I had available for real copper.  Then I used artificial values for the thermal parameters so that my pseudo copper exhibited a rudimentary form of ASB.  I will do the same thing here in a more realistic, position-dependent, dynamical framework.  In this way, I will present what I believe to be an interesting description of ASB and, in addition, a description of yielding transitions in polycrystalline materials.  

In Sec.\ref{EOM} of this paper, I summarize the equations of motion for the thermodynamic dislocation theory, with emphasis on aspects of it that are especially important for present purposes. In Sec.\ref{Expts}, I describe theoretical experiments in which I harden samples by straining them to various degrees and then compute the ways in which they undergo yielding  transitions and shear banding at high strain rates.  The paper concludes in Sec.\ref{Remarks} with remarks about needs for experimental information.

\section{Equations of Motion}
\label{EOM}

\subsection{Basics}

As in \cite{JSL-16}, consider a strip of polycrystalline material, of width $2\,W$, oriented in the $x$ direction, being driven in simple shear at velocities $V_x$ and $-V_x$ at its top and bottom edges.  The total strain rate is $V_x/W \equiv Q/\tau_0$, where $\tau_0 = 10^{-12} s$ is a characteristic microscopic time scale. In contrast to \cite{JSL-16}, here we will look at spatial variations in the $y$ direction, perpendicular to the $x$ axis.  Eventually, we will need to consider general three-dimensional variations in order to model the effects of notches or other crack-initiating spatial irregularities; but, for the present, this simple geometry provides as large a range of dynamical behaviors as is needed.  It is the same as the geometry used by Manning et al. \cite{MANNINGetal-SHEARBANDS-08} in an analysis of shearbanding in amorphous materials.

The local, elastic plus plastic strain rate is $\dot\epsilon(y) = dv_x/dy$, where $v_x$ is the material velocity in the $x$ direction. This motion is driven by a time dependent, spatially uniform, shear stress $\sigma$. Because this system is undergoing steady-state shear, we can replace the time $t$ by the accumulated total strain, say $\epsilon$, so that $\tau_0\,\partial/\partial t \to Q\,\partial/\partial \epsilon$. Then denote the dimensionless, $y$-dependent plastic strain rate by $q(y,\epsilon) \equiv \tau_0\,\dot\epsilon^{pl}(y,\epsilon)$.

The internal state variables that describe this system are the areal density of dislocations $\rho\equiv b^2 \tilde\rho$ (where $b$ is the length of the Burgers vector), the effective temperature $\tilde\chi$ (in units of a characteristic dislocation energy $e_D$), and the ordinary temperature $\tilde\theta$ (in units of the pinning temperature $T_P = e_P/k_B$, where $e_P$ is the pinning energy defined below).  Note that $\rho$ also may be interpreted as the total length of dislocation lines per unit volume, and that $1/\sqrt{\rho}$ is the average distance between dislocations. All three of these dimensionless quantities, $\tilde\rho$, $\tilde\chi$, and $\tilde\theta$ are functions of $y$ and $\epsilon$. 

\subsection{Depinning Rate}

The central, dislocation-specific ingredient of this analysis is the thermally activated depinning formula for the dimensionless plastic strain rate $q$ as a function of a non-negative stress $\sigma$:  
\begin{equation}
\label{qdef}
q(y,\epsilon) = \sqrt{\tilde\rho} \,\exp\,\Bigl[-\,{1\over \tilde\theta}\,e^{-\sigma/\sigma_T(\tilde\rho)}\Bigr]. 
\end{equation}
As shown in \cite{LBL-10,JSL-15}, this formula is an Orowan relation in which it is assumed that the plastic flow is determined entirely by the rate at which entangled dislocations jump instantaneously between near-neighbor pinning sites.  Here, $\sigma_T(\tilde\rho)= \mu_T\,\sqrt{\tilde\rho}$ is the Taylor stress.  It is equal to the ratio of the range of the pinning forces to the average spacing between dislocations (a strain), multiplied by the shear modulus $\mu$; thus $\mu_T$ is a small fraction of $\mu$, and $\sigma_T$ is a geometrically determined stress, mathematically independent of the strain rate, the temperature, or the effective temperature. The fact that the stress dependence occurs in Eq.~(\ref{qdef}) as a function of the ratio $\sigma/\sigma_T$ is important and, I think, very natural; but the exponential function in which that ratio occurs could be replaced by any smoothly decreasing function without changing the qualitative predictions of this theory.  In the following analysis, we shall see that $e_P$ is large, of the order of eV's, so that $\tilde\theta$ is very small, and $q(y,\epsilon)$ is an extremely rapidly varying function of $\sigma$ and $\tilde\theta$.  This behavior is the key to understanding the banding instability.  

\subsection{Dislocation Density and the Onset of Hardening}

The equation of motion for the scaled dislocation density $\tilde\rho$ describes energy flow. It says that some fraction $\kappa_{\rho}$ of the power delivered to the system by external driving is converted into energy of dislocations, and that that energy is dissipated according to a detailed-balance analysis involving the effective temperature $\tilde\chi$.  This equation is: 
\begin{equation}
\label{rhodot}
{\partial\tilde\rho\over \partial\epsilon} = \kappa_{\rho}\,{\sigma\,q\over \tilde\gamma_D\,Q}\, \Bigl[1 - {\tilde\rho\over \tilde\rho_{ss}(\tilde\chi)}\Bigr],
\end{equation}
where $\tilde\gamma_D = \gamma_D/b^2$ is a dislocation energy per unit volume, and $\gamma_D$ is the more familiar dislocation energy per unit length.  Here, $\tilde\rho_{ss}(\tilde\chi) = e^{- 1/\tilde\chi}$ is the equilibrium value of $\tilde\rho$ at given $\tilde\chi$.

It is important to understand the relation between the various ingredients of this formula and the onset of strain hardening.  That rate is defined to be $\Theta_0 \equiv (1/\mu)\,(\partial \sigma/\partial \epsilon)_{\rm onset}$. It has been known for decades (for example, see \cite{KOCKS-MECKING-03}) that $\Theta_0$ often (but not always) remains a material-specific constant over wide ranges of strain rates and temperatures.  We need to understand a physical basis for this rule in order to know when and how to use it.  See, for example, my analysis of the strain-rate anomaly in \cite{JSL-15}. 

To see why $\Theta_0$ may be a constant, consider the following argument made in \cite{LBL-10}.  Hardening begins when the deformation switches from  elastic to plastic so that $q \cong Q$. In most of the situations discussed in \cite{KOCKS-MECKING-03}, the materials apparently have been prepared in such a way that they are relatively free of dislocations.  That is, the initial dislocation density $\tilde\rho$ is still much smaller than $\tilde\rho_{ss}$, so that the energy-conservation law in Eq.(\ref{rhodot}) has the form $\partial\tilde\rho/\partial \epsilon \cong \kappa_{\rho}^{(0)}\,\sigma/\tilde\gamma_D \cong \kappa_{\rho}^{(0)}\,\mu_T\,\sqrt{\tilde\rho}/\tilde\gamma_D$.  Here, I have assumed that the dislocations are still far enough apart from each other that the stress is well approximated by the simple Taylor formula, $\sigma \cong \sigma_T =\mu_T\,\sqrt{\tilde\rho}$.  I also have used a ``bare'' conversion factor $\kappa_{\rho}^{(0)}$ assumed to be strain-rate independent.  Combining these two relations, we find that $\Theta_0 = \mu_T^2\,\kappa_{\rho}^{(0)}/ 2\,\mu\,\tilde\gamma_D$.  Note that this formula is independent of both $\tilde\rho$ and the strain rate, and also is likely to be independent of temperature because $\tilde\gamma_D$ and the elastic moduli ought to scale thermally in the same way.  

Now return to Eq.(\ref{rhodot}) to evaluate the conversion factor $\kappa_{\rho}$.  To do this, it is useful, for stresses that are not too small or negative, to solve Eq.(\ref{qdef}) to find
\begin{equation}
\label{nudef}
{\sigma\over\sigma_T(\tilde\rho)} \cong \ln\Bigl({1\over \tilde\theta}\Bigr) - \ln\Bigl[\ln\Bigl({\sqrt{\tilde\rho}\over q}\Bigr)\Bigr]\equiv \nu(\tilde\rho,q,\tilde\theta).
\end{equation}
To evaluate $\kappa_{\rho}$, we need to look only near onset, where $q \cong Q$ and $\tilde\rho$ is again appreciably smaller than $\tilde\rho_{ss}$. Because $q$ and $\tilde\rho$ appear only as arguments of a slowly varying double logarithm, we can write $\sigma \cong \nu_0\,\mu_T\,\sqrt{\tilde\rho}$, where $\nu_0 \equiv \nu(\tilde\rho_{ss}, Q, \tilde\theta_0)$, and $\tilde\theta_0$ is the scaled ambient temperature.  Now we can repeat the analysis in the preceding paragraph to find that $\kappa_{\rho} = \kappa_{\rho}^{(0)}/\nu_0^2$.  Finally, Eq.(\ref{rhodot}) can conveniently be rewritten in the form
\begin{equation}
\label{rhodot2}
{\partial\tilde\rho\over \partial\epsilon} = \kappa_1\,{\sigma\,q\over \nu_0^2\,\mu_T\,Q}\, \Bigl[1 - {\tilde\rho\over \tilde\rho_{ss}(\tilde\chi)}\Bigr],
\end{equation}
where
\begin{equation}
\label{Theta0}
\kappa_1 = {2\,\mu\over \mu_T} \Theta_0.
\end{equation} 
Note that the factor $\tilde\gamma_D$ has cancelled out, so that the prefactor $\kappa_1$ in Eq.(\ref{rhodot2}) is completely determined by directly observable quantities.  

\subsection{Thermal Equations}
 
The equation of motion for the scaled effective temperature $\tilde\chi$ is a statement of the first law of thermodynamics for the configurational subsystem. The derivation leading to Eq.(2.20) in \cite{JSL-15} tells us that, in the present notation, this equation is
\begin{equation}
\label{chidot0}
c_{e\!f\!f}\,{\partial \tilde\chi\over\partial\epsilon} = {\sigma\,q\over Q}\left(1-{\tilde\chi\over \tilde\chi_0}\right) - \tilde\gamma_D\,{\partial \tilde\rho\over \partial\epsilon},
\end{equation}
where $c_{e\!f\!f}$ is the effective specific heat; and $\tilde\chi_0 \cong 0.25$ (see \cite{LBL-10}) is the steady-state value of $\tilde\chi$ for strain rates appreciably smaller than inverse atomic relaxation times, i.e. much smaller than $\tau_0^{-1}$.  The last term on the right-hand side of Eq.(\ref{chidot0}) is the rate at which configurational energy is stored in the form of dislocations. In \cite{JSL-16} I assumed this term to be negligible.  I will do the same thing in this paper; but I keep the term here because there are circumstances in which it may be important.  (See \cite{JSL-15}.)  

With the same analysis that led from Eq.(\ref{rhodot}) to Eq.(\ref{rhodot2}), Eq.(\ref{chidot0}) becomes
\begin{equation}
\label{chidot}
{\partial\,\tilde\chi\over \partial\epsilon} = \kappa_2\,{\sigma\,q\over \mu_T\,Q}\,\Bigl[ 1 - {\tilde\chi\over \tilde\chi_0} - {\kappa_3\over \nu_0^2} \Bigl(1 - {\tilde\rho\over \tilde\rho_{ss}(\tilde\chi)}\Bigr)\Bigr];
\end{equation}
where the storage term is the expression proportional to $\kappa_3$ inside the square brackets, with
\begin{equation}
\kappa_3 = {\tilde\gamma_D\over \mu_T}\,\kappa_1.
\end{equation}
The overall, dimensionless factor $\kappa_2$ is inversely proportional to $c_{e\!f\!f}$. Unlike $\kappa_1$, whose value is determined directly from experiment via Eq.(\ref{Theta0}), $\kappa_2$ must be determined on a case to case basis by fitting the data. 

The equation of motion for the scaled, ordinary temperature $\tilde\theta$ is the usual thermal diffusion equation with a source term proportional to the input power.  I assume that, of the three state variables, only $\tilde\theta$ diffuses in the spatial dimension $y$. Thus,
\begin{equation}
\label{thetadot}
{\partial\tilde\theta\over \partial\epsilon} = K\,{\sigma\,q\over Q} + {K_1\over Q}\,{\partial^2 \tilde\theta\over \partial\,y^2} - {K_2\over Q}\,(\tilde\theta - \tilde\theta_0).
\end{equation} 
Here, $K = \beta/ (T_P\,c_p\,\rho_d)$, where $c_p$ is the thermal heat capacity per unit mass, $\rho_d$ is the mass density, and $0< \beta < 1$ is a dimensionless conversion factor. $K_1$ is proportional to the thermal diffusion constant, and $K_2$ is a thermal transport coefficient that assures that the system  remains close to the ambient temperature $\tilde\theta_0 = T_0/T_P$ under slow deformation, i.e. small $Q$.

\subsection{Stress}

It remains to write an equation of motion for the stress $\sigma(\epsilon)$ which, to a very good approximation, should be independent of position $y$ for this model of simple shear.  In some applications of this theory, I have used Eq.(\ref{nudef}) to evaluate $\sigma$.  The problem here is that the arguments of $\nu(\tilde\rho,q,\tilde\theta)$ are strongly dependent on $\epsilon$ and $y$, especially in the neighborhood of a shear band. I start, therefore, with the  local relation  $\dot\sigma = \mu[\dot\epsilon(y) - \dot\epsilon^{pl}(y)]$, which  becomes 
\begin{equation}
{d\sigma\over d \epsilon} = \mu\,\left[{\tau_0\over Q}\,{dv_x\over dy} - {q(y,\epsilon)\over Q}\right].
\end{equation}
One simple strategy is to integrate both sides of this relation over $y$ and divide by $2 W$ to find
\begin{equation}
\label{sigmaeqn1}
{d\sigma\over d \epsilon} = \mu\,\left[1 -\int_{-W}^{+W}\,{dy\over 2W}\,{q(y,\epsilon)\over Q}\right].
\end{equation}

An even simpler strategy for numerical purposes is to replace Eq.(\ref{sigmaeqn1}) by
\begin{equation}
\label{sigmaeqn2}
{\partial\sigma\over \partial\epsilon} = \mu\,\left[1-{q(y,\epsilon)\over Q}\right] + M\,{\partial^2 \sigma\over \partial y^2},
\end{equation}
and to use a large enough value of the ``diffusion constant'' $M$ that $\sigma$ remains constant as a function of $y$. I have used both of these strategies for checking the accuracy of the numerical results shown in what follows.  When using Eq.(\ref{sigmaeqn2}), I have chosen $M = 10^5$. 

\section{Theoretical Experiments}
\label{Expts}

\begin{figure}[here]
\centering \epsfig{width=.5\textwidth,file=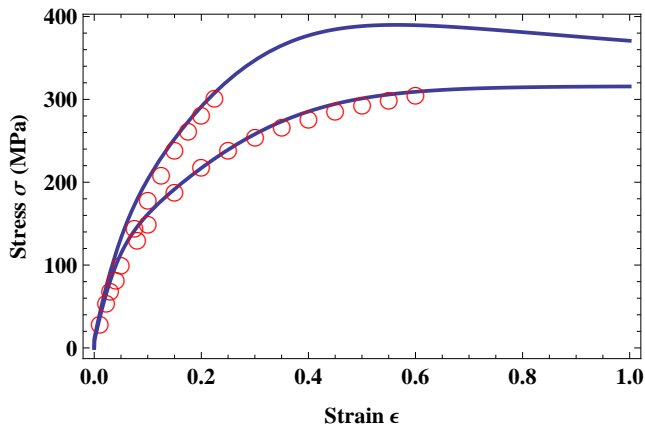} \caption{Hardening curves for $\dot\epsilon = 0.002\,s^{-1}$ (the lower curve) and for $\dot\epsilon = 2,000\,s^{-1}$ (the upper curve).  The red circles are the experimental data used in \cite{LBL-10}} \label{pCuFig1}
\end{figure}
Figure \ref{pCuFig1} shows two room-temperature,  stress-strain curves for real copper, measured and computed at two very different total strain rates, $\dot\epsilon = 0.002\,s^{-1}$ and $2,000\,s^{-1}$.  The experimental points (red circles) are the same as those used in \cite{LBL-10}, where they were taken from \cite{PTW-03,LANL-99}. It is from this data, plus other measurements at other strain rates and temperatures, that my colleagues and I in \cite{LBL-10,JSL-15,JSL-16} obtained values for many of the parameters appearing in the equations written here in Sec.\ref{EOM}.  Specifically, the parameter values to be used in what follows are: $T_P = 40800\,K$, $T_0= 298\,K$, $\mu_T = 1600\,\,{\rm MPa}$,  $\mu = 31\,\mu_T = 39.6\,{\rm GPa}$, $\kappa_1 = 3.1$, $\kappa_2 = 11.2$, and $\kappa_3 = 0$.  

Because I cannot use real copper to study shear banding, I have arbitrarily chosen the thermal coefficients for pseudo copper to be $K = 10^{-5}$ (so that it is slightly smaller than $1/T_P$, i.e. so that the conversion factor $\beta$ is very roughly of the order of unity), $K_1 = 10^{-12}$ (so that thermal diffusion is relevant to the strongly spatial dependent behaviors driving shear banding, but is not so strong as to eliminate those behaviors), and $K_2 = 10^{-9}$ (so as to be roughly comparable in magnitude to the larger values of $Q$, and thus to keep $T \cong T_0$ at smaller strain rates).  The initial values of $\tilde\rho$ and $\tilde\chi$ used for computing both of these curves are $\tilde\rho_i = 10^{-5}$ and $\tilde\chi_i = 0.18$.  Because the thermal terms have not been set to zero for computing the curves in Fig. \ref{pCuFig1}, the upper (fast) curve exhibits thermal softening at large $\epsilon$; but the agreement with experiment at small $\epsilon$ remains quite good. 

Now do the following (theoretical) experiments.  Repeat the slow deformation shown by the lower curve in Fig. \ref{pCuFig1} (for $\dot\epsilon = 0.002\,s^{-1}$), but, this time, stop straining at $\epsilon = 0.2$.  Do this again, for a different sample, stopping at $\epsilon = 0.4$.  Next, make pseudo notches in these pre-strained (i.e. pre-hardened) samples by making spatially localized, negative perturbations of their initial effective temperatures:
\begin{equation}
\tilde\chi(0,y) = \tilde\chi_i - \delta\,e^{- y^2/ 2\,y_0^2},
\end{equation}

\begin{figure}[here]
\centering \epsfig{width=.5\textwidth,file=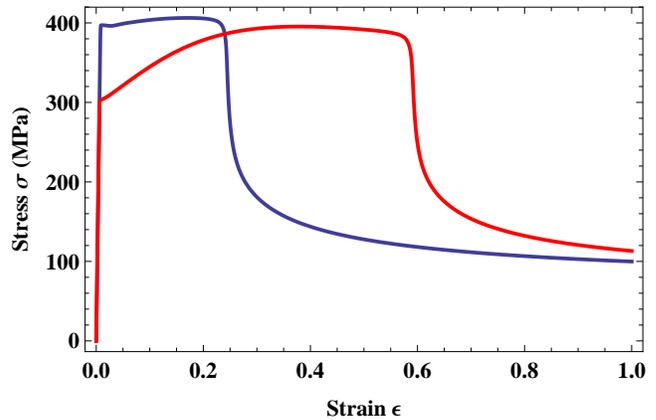} \caption{Stress-strain curves for two pre-hardened samples. The harder sample, shown by the dark curve, fails {\it via} shear banding at the smaller strain. The softer sample, shown by the red curve, fails at a larger strain. } \label{pCuFig2}
\end{figure}

\begin{figure}[here]
\centering \epsfig{width=.5\textwidth,file=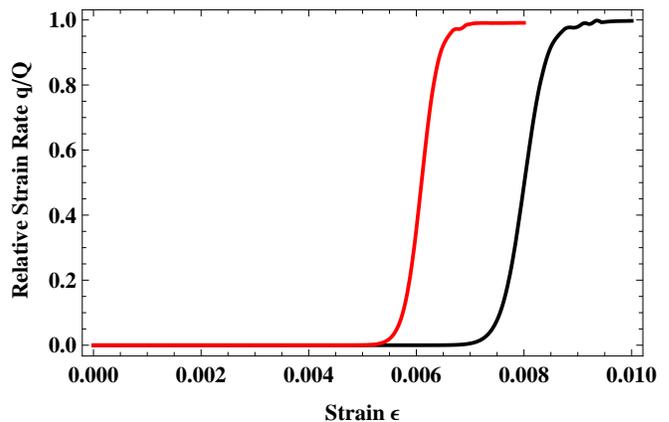} \caption{Relative strain rates $q/Q$ across the initial yielding transition for the two pre-strained samples whose stress-strain curves are shown in Fig.\ref{pCuFig2}. The softer sample, shown by the red curve, is the one that yields earlier.} \label{pCuFig3}
\end{figure}

\noindent with $\delta = 0.02$ and $y_0 = 0.05$.  Finally, strain these samples again at the high rate, $\dot\epsilon = 2,000\,s^{-1}$, by using the final values of $\tilde\rho$ and $\tilde\chi$ in the first  deformations as the initial values for these second stress-strain calculations.  For the first case (the softer, less strained sample), I find these values to be $\tilde\rho_i =  0.0085$, and  $\tilde\chi_i = 0.219$.  For the second case (the harder, more highly strained sample), $\tilde\rho_i = 0.0149$, and $\tilde\chi_i = 0.243$. The resulting stress-strain curves are shown in Fig.\ref{pCuFig2}.  Both samples undergo abrupt stress drops that, as will be seen, indicate shear-banding failures.  The first case, i.e. the harder sample shown by the dark curve in the figure, is the one for which failure occurs earlier; it is the more brittle of the two.  The softer sample, shown by the red curve, fails later; it is tougher.  

\begin{figure}[here]
\centering \epsfig{width=.5\textwidth,file=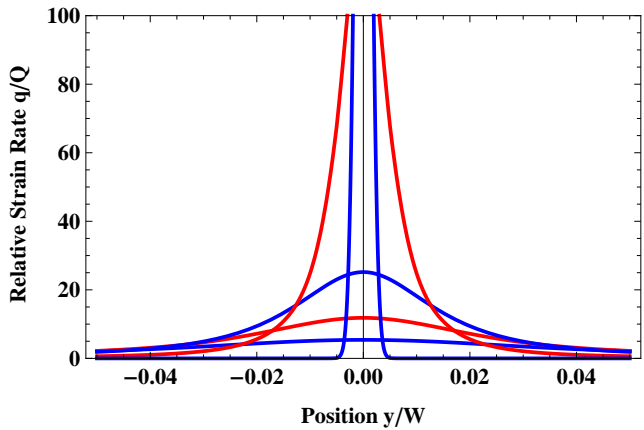} \caption{Relative plastic strain rates $q(\epsilon,y)/Q$ as functions of position $y/W$ for a sequence of increasing total strains $\epsilon = 0.20,\,0.22,\,0.23,\,0.24,\,{\rm and}\,0.25$.} \label{pCuFig4}
\end{figure}

\begin{figure}[here]
\centering \epsfig{width=.5\textwidth,file=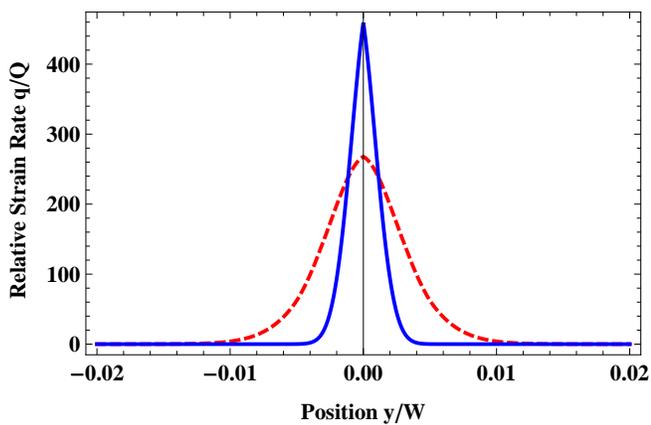} \caption{Relative plastic strain rates $q(\epsilon,y)/Q$ as functions of position $y/W$ for total strains $\epsilon = 0.25,\,{\rm and}\,1.0$. The latter is shown by the dashed line.  Note that, in comparison with Fig.~\ref{pCuFig4}, the horizontal axis has been expanded by a factor of $2$ and the vertical axis compressed by a factor of about $4$.} \label{pCuFig5}
\end{figure}
Before looking in more detail at the shear-banding events, consider what is happening near $\epsilon = 0$, where both samples exhibit what appear to be -- and indeed are -- yielding transitions.  Both curves in Fig.~\ref{pCuFig2} start with very steep elastic sections whose slopes are equal to the shear modulus $\mu = 39.6\,{\rm GPa}$, and then bend sharply to plastic behavior.  These transitions are not infinitely sharp, however. We see in Fig.~\ref{pCuFig3}  that the relative plastic strain rate $q(\epsilon)/Q$ jumps rapidly but smoothly during the transition from elastic to plastic deformation.   (The curves shown here have been computed at $y=0.5\, W$ in order that they not be affected by the pseudo notch at $y = 0$.) The fact that there is a small amount of plastic flow $q$ below the onset point (where $q/Q \to 1$) means that there is a small rate at which dislocations are jumping between pinning sites, consistent with the fact that these systems are known to be noisy near yielding transitions even when plastic flow is unmeasurably small.

Shear band formation near $y=0$ for the harder, more highly strained sample is shown in Fig.~\ref{pCuFig4}.  Plotted here are graphs of $q/Q$ as functions of position $y/W$ for a sequence of increasing total strains $\epsilon = 0.20,\,0.22,\,0.23,\,0.24,\,{\rm and}\,0.25$.  A diffuse shear band is visible at $\epsilon = 0.20$ and becomes increasingly stronger as $\epsilon$ increases.  At $\epsilon = 0.24$, the band is starting visibly to become narrower as it intensifies at the expense of the plastic strain rate at larger values of $y$. Finally, at $\epsilon = 0.25$,  the band has suddenly strengthened and sharpened so much that the strain rate outside this region has  dropped to zero.  Figure \ref{pCuFig5} focusses in on, and expands  this picture vertically, at $\epsilon = 0.25$.  Also shown here by the dashed curve is the plastic flow distribution much later, at $\epsilon = 1.0$.  Apparently, this band has reached its peak intensity and is beginning to spread as heat diffuses away. 

\begin{figure}[here]
\centering \epsfig{width=.5\textwidth,file=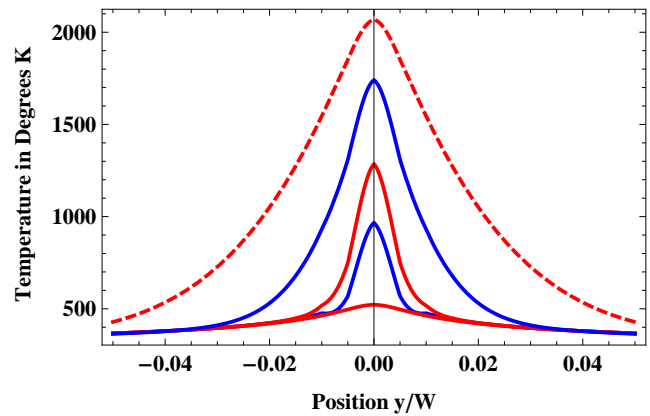} \caption{Temperatures as functions of position $y/W$ at total strains $\epsilon = 0.24,\,0.25,\,0.27,\,0.40,\,{\rm and}\,1.0$, from bottom to top.}. \label{pCuFig6}
\end{figure}

The corresponding sequence of temperature distributions is shown in Fig.~\ref{pCuFig6}.  Here, the sequence of total strains, shown from bottom to top, is $\epsilon = 0.24,\,0.25,\,0.27,\,0.40,\,{\rm and}\,1.0$. Note that the band has achieved its peak sharpness in the strain-rate distribution at the second of these curves, shown in Fig.~\ref{pCuFig5}, for $\epsilon = 0.25$; but it theoretically continues to generate heat for a long time afterwards.  Almost certainly, this behavior is not physically realistic.  At the temperatures shown here, the material inside the band will have melted or undergone other structural changes.  But the onset of rapid failure of one kind or another seems to be a plausible prediction of this analysis.  

While the late stages of the ASB behavior shown in Figs.~\ref{pCuFig4}-\ref{pCuFig6} cannot be realistic in detail, the general picture seems generic for this kind of banding instability.  It is almost identical to the results shown in 1988 by Marchand and Duffy \cite{MARCHAND-DUFFY-88} for shear banding in steel. Within the present set of theoretical experiments, the graphs in Figs.~\ref{pCuFig4}-\ref{pCuFig6} remain almost unchanged when recomputed for the softer sample in  Fig.~\ref{pCuFig2}.  One way to make a bigger change in the banding behavior is to reduce the diffusion constant $K_1$.  Even if I let $K_1 \to 0$, however, the only qualitative change that I see is that the stress drop becomes sharper and deeper, going all the way down to $\sigma \cong 0$, and the band becomes narrow enough to challenge my numerical capabilities.
  
\section{Concluding Remarks}
\label{Remarks}

So far as I know, the microscopic picture of an intrinsically smooth yielding transition presented here is different from the one found in phenomenological descriptions of solid plasticity.  It is also qualitatively different from the picture of yielding in amorphous materials, where transitions between jammed and flowing states are determined by the balance between noise driven creation and annihilation of flow defects, e.g. shear transformation zones (STZ's) \cite{FL-11,JSL-15y}.  Plastic flow in amorphous materials, and their yielding transitions, are determined primarily by their chemical compositions and states of disorder.  These materials do not have long-term memories. 

In polycrystalline solids, however, the flow defects are the dislocations, whose lifetimes are almost infinitely longer than those of STZ's.  These solids do not quickly forget their past deformations. As seen in Sec.\ref{Expts}, the history of a strain hardened sample is partially encoded in its density of dislocations, which determines how it responds to subsequent forcings.  To test this picture, we can observe yielding transitions such as those shown in Fig.~\ref{pCuFig3}.  But, to construct and test a physics-based theory of such transitions, we need independently determined values of parameters like $\mu_T$, $T_P$, $\kappa_1$, etc., for which we need other kinds of experiments. In particular, we need measurements of strain hardening, starting with samples with small dislocation densities; and we need to make those measurements over a range of different temperatures and strain rates.  

The advantage of having detailed material-specific information is that it would allow us to test -- not just the present theory of yielding and shear banding -- but also a wide range of related conjectures.  For example, there is an intriguing set of observations by Rittel and coworkers \cite{RITTEL-08,RITTELetal-10,RITTELetal-12,RITTEL-12} in which they see dynamically recrystallized grains (DRX) appearing in association with, and apparently preceding, the appearance of ASB's.  I would have preferred to write this paper using parameters appropriate for Rittel's ASB-forming titanium alloy instead of using ``pseudo copper.'' Then, discrepancies between my results and the experimental data might have told us whether or not the theory is missing physically essential ingredients.  As I stressed in \cite{JSL-16}, what I have presented here is a bare-bones theory.  It is missing dynamical ingredients such as stacking faults, cellular patterns of dislocations, etc., in addition to DRX.  All of these could be included in the theory in various ways; but we need to find out whether and when to do so in order to draw useful conclusions.

\begin{acknowledgments}

Much of this paper has been written in response to discussions and arguments with D. Rittel, who is therefore responsible for this work even if he strongly disagrees with it.  I thank him greatly. This research  was supported in part by the U.S. Department of Energy, Office of Basic Energy Sciences, Materials Science and Engineering Division, DE-AC05-00OR-22725, through a subcontract from Oak Ridge National Laboratory.

\end{acknowledgments}

\end{document}